\documentclass{ws-ijmpa}
\begin{document}
\markboth{Sujan Sengupta}{Sunyaev-Zel'dovich Effect}

\catchline{}{}{}{}{}

   \title{SUNYAEV-ZEL'DOVICH EFFECT BY MULTIPLE SCATTERING :
NUMERICAL SOLUTION OF THE TRANSFER EQUATIONS}
   \author{Sujan Sengupta \footnote{E-mail : sujan@iiap.ernet.in}}
\address{Indian Institute of Astrophysics,  Koramangala,
              Bangalore 560 034, India} 
   \maketitle

   \begin{abstract}
The radiative transfer equations for multiple
inverse Compton scattering of the
Cosmic Microwave Background Radiation (CMBR) by the hot intra-cluster
electrons are solved numerically. The
spherical isothermal and inhomogeneous $\beta$ model has been
considered for the electron distribution. The
anisotropy of the CMBR caused by scattering, known as thermal Sunyaev-Zel'dovich
effect, along the radial axis of the medium is compared with the
analytical solution of Kompaneets equation. The X-ray data of
several clusters of galaxies at low redshifts provide an estimation
of the central electron density $n_0$ to be of the order $10^{-3}$. It is
found that for this value of $n_0$ the effect of multiple scattering
is negligible.
The numerically calculated anisotropy
along the radial axis matches well with the analytical solution
that describes single scattering. The result incorporating
multiple scattering is fitted with the recent observation of
Sunyaev-Zel'dovich effect in the cluster Abell 2163. 
It is shown that if $n_0$ is greater by
an order of magnitude, which could be possible for cluster of
galaxies at comparatively higher redshift, multiple scattering
would play a significant role at the Wien region of the anisotropy  spectrum.
A fitting formula for the correction to the Sunyaev-Zel'dovich effect due
to multiple scattering is provided.

\keywords{  Cosmology - radiative transfer - scattering - galaxies:
cluster }
   \end{abstract}


\section{Introduction}

Inverse Compton scattering of the Cosmic Microwave Background Radiation (CMBR)
by hot intra-cluster gas - better known as the thermal Sunyaev-Zel'dovich
effect (TSZE)$^{1-3}$
results in a systematic transfer of photons from
the Rayleigh-Jeans to the Wien side of the spectrum causing a distortion
in the Planckian nature of the spectrum. The measurements of the effect yield
directly the properties of the hot intra-cluster gas, the total dynamical
mass of the cluster as well as the indirect information on the cosmological
evolution of the clusters. The effect is also used as an important tool to
determine the Hubble constant $H_{0}$ and the density parameter $\Omega_{0}$
of the Universe$^{4-10}$.
 Recent interferometric imaging of the TSZE$^{11-14}$
has been used to estimate the mass of various galaxy clusters which could
constrain the cosmological parameters of structure formation models.
Measurements of TSZE and the kinetic Sunyaev Zel'dovich effect (KSZE)
from the Sunyaev-Zel'dovich Infrared Experiment and OVRO/BIMA determine the
central Compton y parameter and constrain the radial peculiar velocity
of the clusters. In near future sensitive observations of the effect with
ground based and balloon-borne telescopes, equipped with bolometric multi-frequency
arrays, are expected to yield high quality measurements.

In modeling  the observational data, the analytical
solution of Kompaneets equation$^{15}$ with relativistic corrections$^{16}$
are used
in general. All theoretical discussions and observational inferences so far
are restricted to the case of single Compton scattering.
Single scattering approximation is well justified because X-ray data
of most of the galaxy clusters wherein the Sunyaev-Zel'dovich effect
has been investigated, indicate low optical depth of the intra-cluster
medium (ICM).
 Effects of multiple scattering have been considered by many
authors$^{17-21}$. Recently,
Itoh, Kawana, Nozawa and Kohyama$^{22}$  have studied the effect of
multiple scattering by performing direct numerical integration
of the collision term of the Boltzmann equation and found that
the effect of multiple scattering is negligible for the observed
galaxy clusters.

In this paper, the numerical solutions of the most
general radiative transfer equation for inverse Compton scattering
of CMBR photon are presented by incorporating multiple scattering in a spherically
symmetric inhomogeneous and isothermal medium. Although my approach
is different, the results are in good agreement with that of Itoh
et al.$^{22}$. I further present the results by considering a medium
with optical depth an order of
magnitude higher than that of the galaxy clusters observed so far.

In section~2, I present the most general radiative transfer equation
that describes the inverse Compton multiple scattering of low energy
photons with non-relativistic electrons in an inhomogeneous medium. The
Kompaneets equation which is a special case of isotropic scattering is
also presented  with the analytical solution that describes the
anisotropy in the CMBR spectrum due to single scattering. 
The model parameters are provided in section~3. The results are discussed
in section~4 followed by the conclusion in section~5. 

\section{The Radiative Transfer Equations}

\subsection{General case}

   Chandrasekhar$^{23}$  has provided the radiative transfer equations for
the  Compton scattering by assuming that the electron energy
is much  less than the photon energy. On the other hand, for the
scattering of CMBR, the photon energy is much less than the electron energy.
In fact, in the TSZE, the electrons are considered to have relativistic
motion described by relativistic Maxwellian distribution. The relevant
radiative transfer equation in its general form that describes 
the inverse Compton multiple scattering of low energy photons in a spherically
symmetric inhomogeneous medium can be written as$^{24-26}$
\begin{eqnarray}
\frac{\partial I_{\nu}}{\partial s} &=&\mu\frac{\partial I_{\nu}(\mu,r)}
{\partial r}+\frac{1-\mu^2}{r}\frac{\partial I_{\nu}(\mu,r)}{\partial \mu}
\nonumber \\
&=&-\sigma_Tn_e(r)I_{\nu}(\mu,r) + 
\frac{1}{2}\sigma_Tn_e(r)\omega_0 \times \nonumber \\
& &\int^1_{-1}\left\{P_0-\frac{2kT_e}{m_ec^2}P_1+
\frac{2kT_e}{m_ec^2}\left(\nu^2\frac{\partial^2}{\partial\nu^2}-   
2\nu\frac{\partial }{\partial\nu}\right)P_2\right\}
I_{\nu}(\mu,r)d\mu'.
\end{eqnarray}

The above equation describes the inverse Compton scattering of
low-frequency radiation (first order in $h\nu/mc^2$) in a hot
thermal isotropic electron gas (first order in $kT_e/mc^2$) in the
rest frame of the electron gas.
Here $\omega_0$ is the albedo for single scattering, $s$ is the ray path, 
$\mu=\cos\theta$ where $\theta$ is the angle between the axis of symmetry
(radial axis) in the rest frame of electron gas and the ray path,
$\sigma_T$ and $n_e$ being the Thomson scattering cross section and the electron
number density respectively.
 $k$, $c$, $m_e$, $\nu$ and
$T_e$ are Boltzmann constant, velocity of light, electron rest mass, frequency
of the photon  and the temperature of the electron respectively.
 $\omega_0=1$ for a purely
scattering medium. The phase functions $P_0$, $P_1$ and $P_2$ are given as :
\begin{eqnarray}
P_0(\mu,\mu')=\frac{3}{8}\left[3-\mu^2-\mu'^2(1-3\mu^2)\right],
\end{eqnarray}
\begin{eqnarray}
P_1(\mu,\mu')=\frac{3}{8}\left[1-3\mu'^2-3\mu^2(1-3\mu'^2)+2\mu^3\mu'(3-5\mu'^2)
+ 2\mu\mu'(3\mu'^2-1)\right],
\end{eqnarray}
and
\begin{eqnarray}
P_2(\mu,\mu')=\frac{3}{8}\left[3-\mu^2-\mu'^2+\mu\mu'(3\mu\mu'-5+3\mu^2+3\mu'^2-
5\mu^2\mu'^2)\right].
\end{eqnarray}

\subsection{Special case : Isotropic radiation field}

For an isotropic radiation field
\begin{eqnarray}
 \frac{1}{2}\int_{-1}^{1}{I(\mu',\nu)d\mu'}=I(\nu),
\end{eqnarray}
\begin{eqnarray}
\frac{1}{2}\int^1_{-1}{I(\mu',\nu)\mu'^2d\mu'}=\frac{1}{3}I(\nu) 
\end{eqnarray}
and
\begin{eqnarray}
\int^1_{-1}{I(\mu',\nu)\mu'd\mu'}=\int^1_{-1}{I(\mu,\nu)\mu'^3d\mu'}=0.
\end{eqnarray}

Therefore, for isotropic scattering, equation (1) reduces to
\begin{eqnarray}
\frac{\partial I_{\nu}}{\partial s}=
\sigma_T n_e(r)\frac{2kT_e}{m_ec^2}\left(\nu^2\frac{\partial^2I_{\nu}}{\partial\nu^2}-2\nu\frac{\partial I_{\nu}
}{\partial\nu}\right)
\end{eqnarray}
which is well known as the Kompaneets equation for low frequency photons.

Integration of equation (8) along the ray path provides the thermal component of
the distortion $\Delta I$ and is written as
\begin{eqnarray}
\Delta I(x)= I_0 y\frac{x^4e^x}{(e^x-1)^2}\left[\frac{x(e^x+1)}{e^x-1}-4\right]
(1+\delta_T)
\end{eqnarray}
where $x=\frac{h\nu}{kT_{CMBR}}$, $I_0=2(kT_{CMBR})^3/(hc)^2$.
 The term  $y=\frac{2kT_e}{m_ec^2}\int{\sigma_Tn_eds}$ 
is usually referred to as the Compton y parameter and $\delta_T$ is a relativistic
correction to the thermal effect$^{16}$
significant if $kT_e > 5 $KeV.

\section {The Models}

I have adopted the isothermal $\beta$ model$^{27}$
for the density distribution of the clusters. The spherical
model density is described by
\begin{eqnarray}
n_e(r)=n_0\left(1+\frac{r^2}{r_c^2}\right)^{-3\beta/2},
\end{eqnarray}
where the core radius $r_c$ and $\beta$ are shape parameters,
$n_0$ is the central electron density.

I have adopted the values of the model parameters given in LaRoque et al.$^{13}$
for
the Sunyaev-Zel'dovich anisotropy spectrum observed in Abell 2163.
A combination of ROSAT X-ray data and OVRO/BIMA observation of
Sunyaev-Zel'dovich effect of Abell 2163 at the cosmological redshift z=0.203
implies the electron temperature $T_e=12.4$ KeV, the angular core
radius of the cluster $\theta_c=1'.20\pm 0'.11$ and the shape
parameter $\beta=0.616\pm 0.031$. However, for simplicity, 
I have adopted $\beta=2/3$ which also provides a good
fit to the observed data. Following LaRoque et al.$^{13}$ , I have
considered a cosmological model with the matter density parameter
$\Omega_m=0.3$ and the cosmological constant corresponding to
$\Omega_v=0.7$. I have taken the Hubble constant $H_0= 58.0$ Kms$^{-1}$Mpc$^{-1}$. 
With the above parameters, observational fit of the Sunyaev-Zel'dovich
effect for Abell 2163 yields the central electron number density
$n_0=6.96\times 10^{-3}$ and hence the corresponding Compton y parameter is 
$3.83\times 10^{-4}$. In order to investigate the effect of multiple scattering
in a denser medium I have also considered models with $10\times n_0$ and
$20\times n_0$ keeping other parameters unaltered.

\section{Results and Discussion}

Equation (1) is a coupled integro partial differential equation and so
it cannot be solved analytically. I have solved it numerically by discretization
method. In this method the medium is divided into several shells and the
integration is performed over two dimensional grids of angular and radial
points. The numerical method is described in detail in
Peraiah \& Varghese$^{28}$. I have used 11 points Trapezoidal method for angular
and frequency integration. The numerical code is thoroughly tested for stability and
flux conservation. 

For the boundary condition, I have provided equal amount of intensity
corresponding to $T_{CMBR}=2.728$K at optical depth $\tau=0$ along all directions, i.e.,
\begin{eqnarray}
I(\pm \mu,\nu,\tau=0)= \frac{2h\nu^3}{c^2}\left(e^{h\nu/kT_{CMBR}}
-1\right)^{-1}.
\end{eqnarray}
The relativistic correction $(1+\sigma_T)$ is included to the
emergent intensity.

Equation (1) decribes the anisotropic and inhomogeneous radiation
field in a spherically symmetric medium. Although scattering makes the
radiation anisotropic, spherical symmetry of the medium and the initial
isotropy of the radiation make the emergent radiation field isotropic.  As 
a consequence the distortion along the radial axis and that along the
ray path is the same.

\begin{figure}
\centerline{\psfig{file=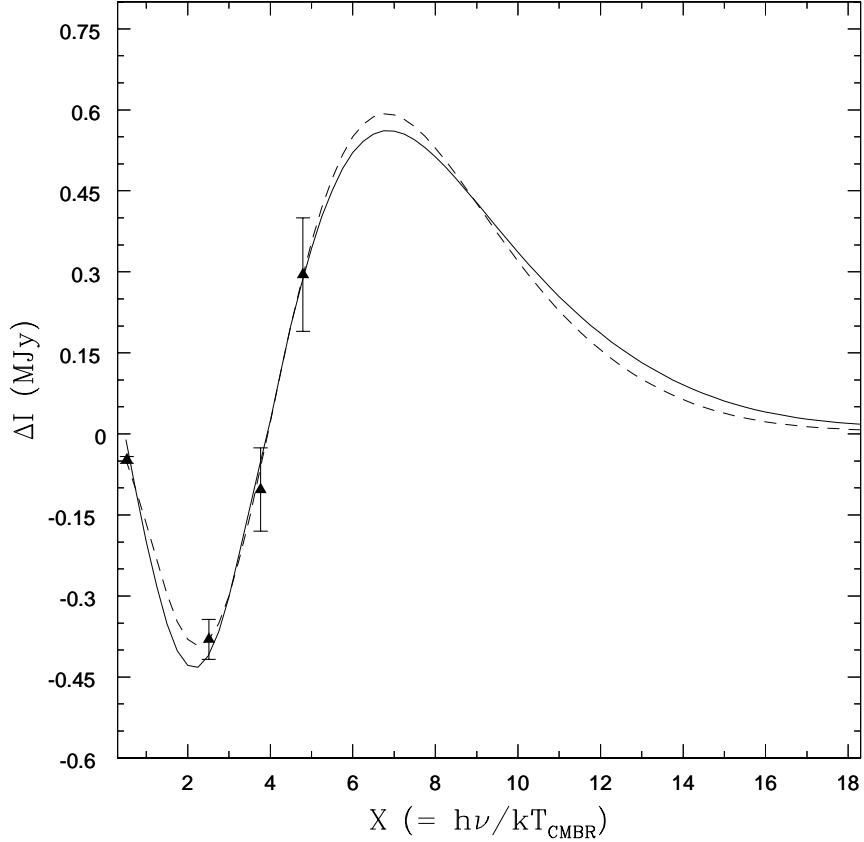,width=12cm}}
\vspace*{8pt}
\caption{ Best model fit for thermal Sunyaev-Zel'dovich spectrum of Abell 2163. 
Solid line represents the distortion along the radial axis for multiple
scattering and dashed line represents that along the ray path for single
scattering. The central electron number density $n_0=6.96\times 10^{-3}$}
\end{figure}

The numerical results that incorporate multiple scattering and the results
for single scattering are presented in Fig~1, Fig~2 and Fig~3 for different
values of the electron number density.
Since the optical depth with a central electron number density $n_0=
6.96\times 10^{-3}$ is small, the effect of multiple scattering is negligible
as seen in Fig~1.
The slight
differences in the results at the Wien and Rayleiegh Jeans region
is well in agreement with that obtained by Itoh et al.$^{22}$ . This can be
visualized from Fig.~1 where I have fitted the observational data for
the galaxy cluster Abell 2163. The model fit clearly shows
that the effect of multiple scattering can be neglected for Abell 2163
within the uncertainties in the physical parameters.

\begin{figure}
\centerline{\psfig{file=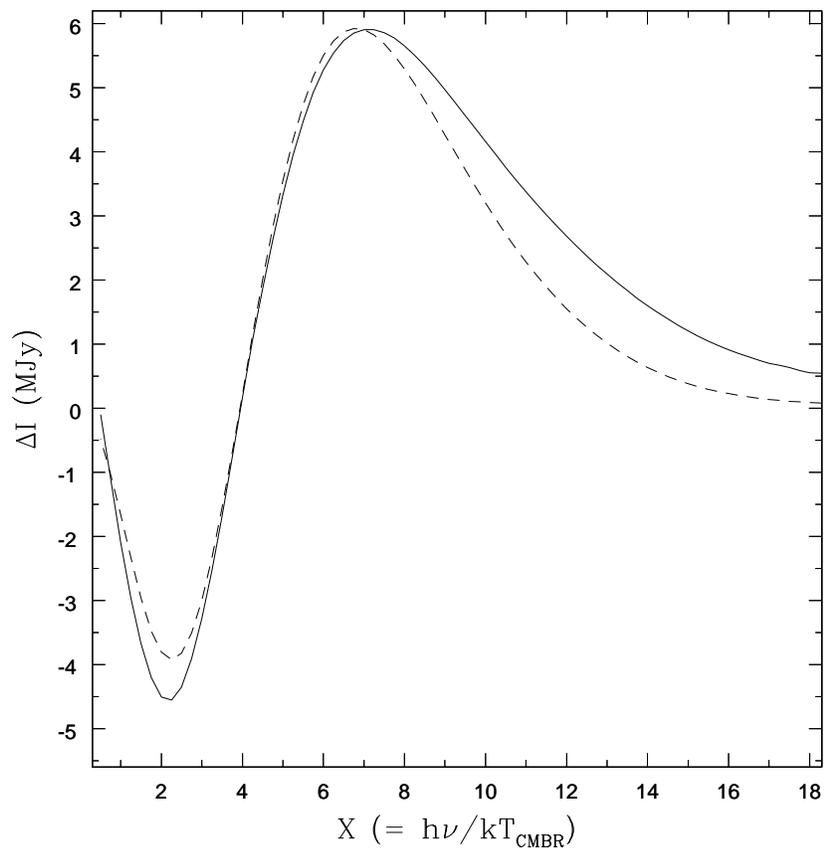,width=12cm}}
\vspace*{8pt}
\caption{Same as Fig.~1 but with the central electron density 
$10\times n_0$.}
\end{figure}
 
\begin{figure}
\centerline{\psfig{file=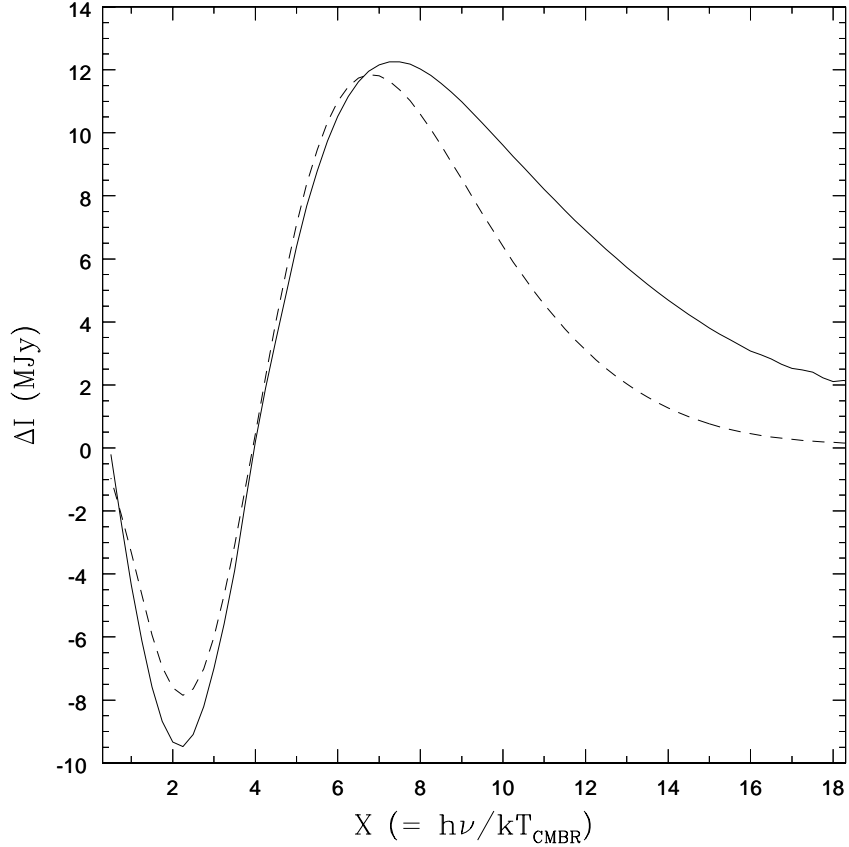,width=12cm}}
\vspace*{8pt}
\caption{ Same as Fig.~1 but with the central electron density 
$20\times n_0$.}
\end{figure}
 
In Fig.~2 and Fig.~3 the effect of multiple scattering is shown by increasing
the central electron density ten and twenty times respectively. The other
parameters such as the electron temperature, shape parameters etc. and the cosmological
parameters are kept unchanged. With the increase in the electron number
density and hence the optical depth of the medium, the number of
scattering increases. Consequently, the distortion in the spectrum increases
significantly. The increase in the distortion is much more at the Wien
region of the spectrum than that at the Rayleigh-Jeans region.

The spectral distortion due to Sunyaev-Zel'dovich effect by single
scattering is characterized by three distinct frequencies : the cross over
frequency $x_0=3.83$ where the thermal Sunyaev Zel'dovich effect
vanishes, $x_{min}=2.26$ which gives the minimum decrement of the
CMB intensity and $x_{max}=6.51$ which gives the maximum distortion.
The value of $x_0$ is pushed to higher values of $x$ with the increase
in $T_e$ for the relativistic case. Fig.~2 and Fig.~3 shows that there
is no chnage in the values of $x_0$, $x_{min}$ and a negligible change in
$x_{max}$ when multiple
scattering is included. The amount of distortion due to multiple scattering
is however higher at $x_{min}$ as compared to that with single scattering.
At $x_{max}$ the amount of distortion remains almost the same even if the
central electron density is increased by twenty times. The effect of multiple
scattering becomes significant as $x$ increases from $x_{max}$ that is at the far Wien
region. Therefore, multiple scattering must be incorporated in
modeling the thermal Sunyaev-Zel'dovich effect anisotropy spectrum
if the optical depth of the medium is comparatively high. Quantitatively,
if the central electron density of a galaxy cluster is one order more
than that of Abell 2163 or other clusters observed so far then the effect of
multiple scattering would be very important.  

Finally, the thermal component of the distortion $\Delta I$ modified due
to the multiple scattering can be written as :
\begin{eqnarray}
\Delta I(x)= I_0 y\frac{x^4e^x}{(e^x-1)^2}\left[\frac{x(e^x+1)}{e^x-1}-4\right]
(1+\delta_T)(1+\delta_M)
\end{eqnarray}
where the correction due to multiple scattering $\delta_M$ is expressed
by a fitting formula given by
\begin{eqnarray}
\delta_M&=&7.2n_0\left[1+\log\left(\frac{0.139}{n_0}\right)\right]
      (-1.864897+2.744115x-1.337648x^2+ 
      0.307672x^3 \nonumber \\
      & & -3.812115\times10^{-2}x^4+2.634191\times10^{-3}x^5-
       9.551485\times10^{-5}x^6+ \nonumber \\
      & & 1.432487\times10^{-6}x^7).
\end{eqnarray}

\section{Conclusions}

The general transfer equations for 
inhomogeneous low frequency radiation multiply scattered by relativistic
thermal electron gas in a spherically symmetric medium are solved
numerically. 
The observed Sunyaev-Zeldovich effect for Abell 2163 is fitted with the
numerical result. The numerical result for the
distortion in the CMBR due to multiple scattering is compared
with that given by the analytical solution of
Kompaneets equation. The effect of multiple scattering is found
to be insignificant for the inferred optical
depth of the galaxy clusters observed so far. However, if the central electron number
density is increased by an order of magnitude, multiple scattering
affects the CMBR distortion spectrum significantly in the Wien region.
Therefore, it is important to include multiple scattering while
modeling the observed Sunyaev-Zel'dovich effect of galaxy clusters with
high electron density. A fitting formula for incorporating the correction
due to multiple scattering is provided.

\section*{Acknowledgments}

I am thankful to the referee for useful suggestions. Thanks are due to
P. Bhattacharjee for discussions.

\end{document}